\documentstyle[12pt,aasms4]{article}
\begin{document}
 
\title{A 106-d period in the nuclear source X-8 in M33}

\author{Guillaume Dubus\altaffilmark{1} and Philip A. Charles}
\affil{University of Oxford, Dept. of Astrophysics, Keble Road, Oxford OX1~3RH, UK}
\author{Knox~S. Long}
\affil{Space Telescope Science Institute, 3700 San Martin Dr., Baltimore MD 21218}
\author{Pasi~J. Hakala}
\affil{Mullard Space Science Laboratory, UCL, Holmbury St. Mary, Dorking, Surrey RH5~6NT, UK}
\altaffiltext{1}{Present address: DARC, UPR 176 du CNRS, Observatoire de Paris Meudon, 5 place Janssen, 92195 Meudon, France}

\begin{abstract}
With an X-ray luminosity of about $10^{39}$\rm erg$s^{-1}$, the source X-8
coincident with the optical center of M33 is the most luminous X-ray source
in the Local Group. However, its nature remains a mystery. 
We present
here new and archival ROSAT observations of X-8 spread over 6 years which
show variability and a
$\sim$ 106-d periodicity.
This implies that (most of) the emission from M33 X-8 arises 
from a single object, perhaps 
a binary system with a $\sim$ 10 M$_{\odot}$ black hole
primary.  
We suggest that
the companion is a giant orbiting with a $\sim$ 10-d period,
and that the observed modulation is ``super-orbital'', analogous to that
seen in Cyg X-2 and X1820-30.  

\end{abstract}


\section{Introduction}

The nearby spiral galaxy M33 was first detected in X-rays using 
the {\it Einstein
Observatory} Imaging Proportional Counter (IPC, \cite{long}) and the High
Resolution Imager (HRI, \cite{markert}; \cite{trinchieri}). These
observations revealed 
a bright source, X-8, coincident with M33's faint optical
nucleus.  
The source comprised almost 70\% of the total X-ray luminosity of M33.
At M33's distance, 795 kpc (\cite{vand}), X-8 has
a (0.15--4.5 keV) $L_{X}$ of $\sim10^{39}\rm  \:erg\:s^{-1}$ 
and is the brightest source in the
Local Group. Other observations of X-8 have been made
using {\it EXOSAT} (\cite{gott}), {\it ASCA} (\cite{takano}), and the {\it
ROSAT} Position Sensitive Proportional Counter (PSPC, \cite{long2}) and HRI
(\cite{schulman}). However, the nature of
this source is still unresolved.  Possibilities include a quiescent mini-AGN
(\cite{trinchieri}; \cite{peres}), a collection of X-ray binaries (\cite{hernquist},
hereafter HHK) and a new type of X-ray binary (\cite{gott}).

The temporal properties of X-8 are likely to be crucial to unraveling
the problem.  However, not much has been known other than that the source
was persistent at about the same luminosity over a 15 year baseline.
Markert \& Rallis (1983) did find (from {\it Einstein} HRI
data) that the flux from X-8 had decreased by 40\% on a time scale of six
months, but similar variations were not detected in nearly simultaneous IPC and MPC observations.  
As a result, Peres et al. (1989) argued that the
variability only affected energies below 1.2 keV. To address this
problem, we have conducted a new study of the X-ray
variability of X-8 based on {\it ROSAT} PSPC and HRI observations made between
July 1991 and January 1997. Our results concerning the other bright sources
will be published elsewhere.

\section{Observations}

Multiple {\it ROSAT} HRI and PSPC (0.1-2.4 keV) 
observations of M33 have been carried
out (See Tr\"umper, 1984, for instrumental
details). The majority of these were centered on the optical nucleus of M33
at $\alpha~1^{\rm{h}}{33}^{\rm{m}}50\fs4$,
$\delta~30^\circ39\arcmin36\arcsec$ (J2000), and even in the
off-center pointings X-8 was in the field of view.
In that period, the 
HRI and PSPC exposures totalled 387 ks and 62 ks respectively, of which 233 ks
and 50 ks were centered on X-8
(see table 1).
Some of these HRI observations, {\it rh20a} and {\it b}, were analyzed by  
Schulman and Bregman (1995) and some of the PSPC observations,  
{\it rp23a}, {\it b} and {\it c},
were described by Long et al. (1996),
but the remainder have not been discussed previously.  X-8 appears 
as a point source in all of the images.  The point source is
superposed on low surface brightness diffuse emission extending 
throughout the region around the nucleus, as has been described by 
Schulman \& Bregman (1995). 

To study the X-ray properties of X-8, we extracted counts
from a 1\arcmin~aperture 
centered on X-8 and estimated
the background from a source-free annulus (with inner and outer radii of
2\farcm 8 and 4\arcmin) centered on X-8. The HRI data were extracted 
initially as barycenter-corrected photon arrival times, subsequently
binned per orbit (defined as the average flux in good time intervals
within 3000 s of each other, the average time when M33 is
continuously viewable by {\it ROSAT}). The rates were then corrected for
vignetting and background using the HRI point spread function (David {\it et
al}, 1996).  
This made it possible to compare centered and off-centered fluxes. 
(The vignetting correction for the HRI was in fact quite small,
$<$5\%.)  
A similar technique was used for the PSPC data, but we only
used the three off-axis PSPC observations individually, as correction
uncertainties made comparisons between them impossible.
The average corrected HRI and PSPC count rates were about 205 ks$^{-1}$ and 
580 ks$^{-1}$ (compared to background rates of 15 ks$^{-1}$ and 5
ks$^{-1}$), respectively.  

\section{Analysis and Results}

We processed the data in three different formats: 
event lists, corrected mean orbital fluxes, and corrected mean
observation fluxes. The photon arrival times were tested for variability 
using the Kolmogorov--Smirnov (KS) and
the Cramer--Smirnov--von Mises (CSM) tests. We searched for
power at frequencies between $10^{-3}$ and $10$ Hz using a modified Fourier
spectrum (\cite{dubus2}) and for variability time scales following Collura
et al. (1987). On corrected mean fluxes, we applied the
standard $\chi^2$ test and the Lomb-Scargle (LS)
normalized periodogram. We also tested the corrected mean fluxes for
variability using the method of
Maccacaro et al. (1987).

The analyses were applied independently to each observation and to
various combined datasets : {\it rh8} combines {\it rh83} to
{\it rh60a} (taken over 2 contiguous weeks), {\it cen-h} combines
all centered ($\leq$~1\arcmin offset) HRI observations,
{\it hri} has all the HRI data, and {\it cen-p}
contains all centered PSPC observations. The data were combined both as mean
orbital fluxes and observation averages, except for {\it rh8} where only
orbital means were combined. We did not attempt
to include the 10-ks off-centered PSPC data into a single set nor to
combine HRI and PSPC data. Tests on (uncorrected) arrival times were not
applied to data combining on and off-centered data.

Within single orbits, the KS and CSM unbinned tests indicated time
variability in three of the PSPC observations. 
Further analysis using the technique described by Dubus et al. (1998) revealed
significant power at period of about 400 s. Thus, the short term
variability that we observed is due to the 400 s `wobble' used
standardly during {\it ROSAT} observations to reduce spatial variations
in the detector response. As might be expected, in observation
{\it rp23c} when the amount of `wobble' was reduced, there was no
significant power at 400 s.  Averaging the flux on spacecraft
orbits eliminates this effect. 
Significant peaks were not found in power spectra of
the on-axis HRI data, with an upper limit of 20\% (15\% with the on-axis
PSPC data) on any sinusoidal modulation at these frequencies
(\cite{dubus2}).  Hence, the {\it ROSAT} data did not reveal any
short term variability in X-8.

Within individual observations, X-8 is also fairly constant.
The only exceptions were for {\it rp23c}
when the flux increased by $\ga$ 10\% in a few hours and 
{\it
rh11a} when the flux may have decreased by $\ga$ 10\% within a few days.
The variations in count rate in {\it rp23c} were seen both in 
the soft ($<$ 0.84 keV) and hard ($>$ 0.84 keV) bands.
This was also true if the boundary between soft and hard was set at 1.2
keV.

X-8 does appear to be variable in {\it rh8} (which combines {\it rh83}--
{\it rh60a}).  This is the best sampled time interval with an exposure
time 163 ks time over 2 weeks (figure \ref{fig:rh8}). 
We believe that the erratic behavior of X-8 during this time interval
(figure \ref{fig:rh8}) is real:
(1) The on-axis observation is consistent with this hypotheses. 
(2) The background represents at most 10\% of the raw flux while 
there are $\ga$ 20\% variations. (3) Removing observations
{\it rh85} and {\it rh87} (the furthest off-axis) or (4) changing the
extraction regions has no effect.  And (5) similar variations were
observed neither 
in the background nor in any other bright source in the field of view.

When observations over long time intervals ($\ga$ 1 week) are considered, 
variability becomes far more obvious.  As shown in
figure \ref{fig:avg}, 
X-8 varies by $\ga$ 20\% on a
timescales of months.  There are no reports of changes in the
instrument sensitivity that could explain such variations. 
The
PSPC data are included by assuming that the mean fluxes from the temporally
close (within 10 days, see table 1) observations {\it rp232} and
{\it rh20b} were the same and renormalized accordingly. PSPC and HRI data
were {\it not} grouped together for analysis. 

When a LS analysis is carried out on the long datasets, a 
106-d period is found with very high confidence in the HRI data.
The LS analysis was confirmed with other methods contained in
the STARLINK PERIOD package: CLEAN,  $\chi^2$ fit
to a sine wave and phase dispersion minimization (PDM) methods. All
yield a period of 105.9 $\pm$ 0.1 days.  
The data folded on the 106-d period are shown in figure \ref{fig:period}.
With this period, the erratic behavior 
in dataset {\it rh8} can be associated with the rise time at phases 0.8-1.0.
The folded renormalized PSPC
data fits in very well, although it does not separately reveal 
a 106-d periodicity, due, at least in part, to the fact that there
are so few PSPC observations.  

The period is also found in the centered observations alone, whether
background-corrected or not, and as such, is not an artifact of the
corrections and/or pointings. Removing each observation in turn to
see if the detection is due to one in particular, we find either a $\sim$
52-d or a 105.9-d peak whenever there is a sufficient time base. The alias
at $\sim$ 52 days is ruled out by the last HRI observation {\it rh03}
(obtained at phase 0.40-0.45, figure \ref{fig:period}).
The associated background showed no structure when folded on the 106-d
period, and exhibits no variation greater than 10\% of the amplitude of the
X-8 modulation.

Simulating the X-8 data with a constant Poisson flux at the same mean gave
no significant peaks in the LS analysis.  Also, colored noise is unlikely
at such low frequencies.  While active galactic nuclei have a red noise
power spectrum above $\sim 10^{-5}$ Hz, white noise dominates below that
(\cite{hardy}). The robustness to different changes in
the data, the amplitude of the signal, the good phase coverage and the fit
of the PSPC data all testify in favor of the reality of this periodic
behavior.

For completeness, figure \ref{fig:period} also contains
the folded mean fluxes of the 3 {\it Einstein} HRI and 2 IPC observations
as given by Peres et al. (1989). (Details of the cross calibration between
the {\it Einstein} HRI and IPC can also be found there.) 
Our renormalisation to the {\it ROSAT} HRI count rate is arbitrary. 
Despite the normalization uncertainties, the
{\it Einstein} HRI data appear consistent with the 106-d period. 
One of the IPC points may be discrepant, but with only two measurements
this discrepancy could easily be due to the difficulties associated
with normalization.  There were simply not enough {\it Einstein} observations
to obtain a robust result.

\section{Discussion}

Although X-8 is the brightest X-ray source in the Local Group, the nucleus
of M33 is optically inconspicuous and semi-stellar
(Kormendy and McClure,1993, hereafter KM), with a core
radius $\le$ 0.3 pc and a low velocity dispersion of about 21 km s$^{-1}$.
This implies a central relaxation time $\sim 10^7$ years. Thus, it is
likely that the nucleus has undergone core collapse. HHK suggested
that the M33 nucleus resembles a globular cluster and that the intense X-ray
emission is due to $\sim$ 10 low mass X-ray binaries (LMXBs) formed during
the core collapse, each with a luminosity of $\sim 10^{38}$ ergs$^{-1}$.
(As noted by HHK, this requires sources 
brighter by a factor ten than galactic globular
cluster LMXB sources).

A 106-d period renders this scenario unlikely.
A 10\% amplitude for the X-8 periodic variations requires
that one of the ten objects modulates its luminosity by almost 100\%, i.e. that
it is transient. But this poses several problems: (a) This would be by far
the shortest known transient recurrence time; (b) the quiescence interval
would be extremely short; and (c) the outburst regularity would be exceptional.
X-ray transients such as the neutron star LMXB Aql X-1 or the black hole
candidate 4U1630-47 show outbursts on {\it timescales} of a few hundred days
but are not  periodic (\cite{kuulkers}). The detection
of the 106-d period over $\sim$20 cycles implies a regularity and a short
duty cycle incompatible with typical soft X-ray transients.
We conclude that most of the X-8 luminosity
arises from a single object.

One possibility is that the M33 nucleus contains a quiescent AGN.
However, KM's upper limit of $10^4$~M$_{\odot}$ on the mass 
contained in the inner 0.3 pc of M33 rules out the
presence of a supermassive black hole and makes interpretation of X-8
as a quiescent AGN untenable.
This conclusion is further supported by the {\it ASCA} (\cite{takano}) and
{\it EXOSAT} (\cite{gott}). X-ray spectra of X-8, which are best described by
a power law plus an exponential cutoff above 2 keV, unlike known
AGN. 

On the other hand, core collapse could have lead to the formation of 
a stellar mass black hole (KM),
with L$_X$ $\rm \sim 10^{39}$~ergs$^{-1}$ corresponding to an Eddington 
limited 10
M$_{\odot}$ black hole accreting at
$\dot{M} \sim 10^{-7}$~M$_{\odot}$yr$^{-1}$. 
The X-8 spectrum is softer than typical neutron star LMXBs, but
comparable to LMXB black hole candidate spectra (\cite{takano}). 

In principle, the mass transfer could arise from the wind 
of a massive O--B star. But such
stars are not expected in the globular cluster-like nucleus of M33 and hence
are excluded.
Alternatively, tidal capture in the core could form a system in which the
companion is degenerate. In this case, $\dot{M}$ implies $P_{\rm orb}\sim
0.1$hr and the mass of the donor star would be $\sim0.1$ M$_{\odot}$
(\cite{king}), similar to the globular cluster source X1820-30. This
neutron star LMXB has a $\sim$ 20\%, 176d ``super-orbital'' variation, the
origin of which is unknown. However, as the time 
scale for angular momentum losses through gravitational radiation is 
close to $10^4$yr ($\sim 10^6$yr for 
X1820-30), it is unlikely that we are observing a degenerate system in X-8.

A 106-d orbital period would be compatible with an evolved companion with 
a 0.3 M$_{\odot}$ helium core and a total mass of $\sim$ 2 M$_\odot$ 
(\cite{king2}). This would make it the longest known orbital period in an
X-ray binary.
But King et al. (1997a,b)
have shown that such systems would certainly be transient and this
is not observed here. To make the source persistent, one is led to 
orbital 
periods of $\sim$ 10d and inconsistent mass ratios close to unity.

The assumptions of King et al.
(1997a,b) do not hold if the companion is massive enough. For instance,
the transient GRO J1655-40 is a 7 M$_{\odot}$ black hole binary
with a 2.3 M$_{\odot}$ companion and $\dot{M}\sim$10$^{-7}$ M$_{\odot}$
yr$^{-1}$ (\cite{orosz}). The
companion is probably crossing the Hertzprung gap, evolving to the giant
branch, and this places GRO J1655-40 in a narrow transient strip in an
otherwise persistent luminosity domain (\cite{kolb}). Following figure 2 of
Kolb et al. (1997), it is thus conceivable
to have a persistent source such as X-8 if the companion is a $\ga 2.5$
M$_{\odot}$ giant with $P_{\rm orb}\la 10$-d.  We note that
O'Connell (1983) found evidence in the nucleus of M33 for an increased
population of intermediate mass (2-5 M$_{\odot}$) stars as compared to the
M31 nucleus or our galaxy.

The origin of the 106-d period is still a problem. Precession of a 
warped disk (\cite{maloney}) would lead to eclipses rather than a modulation.
King et al. (1997a) used the van Paradijs (1996) criterion for
instability, i.e. that a necessary condition for instability is that the
disk temperature be somewhere in the disk below the hydrogen ionization
temperature of 6500 K. The condition for stability is most stringent at the
outer radius where the temperature is lowest. Irradiation of the disk by the
central X-ray source can have a major influence on this temperature
(\cite{king2}). With a $\sim$ 2.5 M$_{\odot}$ giant companion and $P_{\rm
orb}\sim 10$-d, the outer disk radius would be much larger than in most
binary systems and irradiation could be limited only to the inner parts of
the accretion disk. Relaxing the van Paradijs criterion could allow
instabilities to exist in the outer regions, leading to variability in
the source, and possibly ``super-orbital'' 
modulation. This will be the subject of a future paper.

\section*{Acknowledgments}
We thank Jean-Pierre Lasota, Jean-Marie Hameury, Andrew King and
Erik Kuulkers for fruitful discussions on various parts of this work. 
We acknowledge support by the British-French joint research
program {\it Alliance} and by NASA grant NAG 5-1539 to the STScI.
This research has made use of data obtained through the High Energy
Astrophysics Science Archive Research Center Online Service, provided by the
NASA-Goddard Space Flight Center.

\pagebreak

\begin{table} 
\begin{center}
\begin{small}
\begin{tabular}[h]{lclc}
\multicolumn{4}{c}{Table 1: {\it ROSAT} observations of M33 
}\\
\hline
\hline
Obs.\tablenotemark{a} & Offset & Dates & Duration\\
 & (\arcmin) & & (ks)\\
\hline
\sc{pspc} \hfill \\
\hline
$rp23a$ &  0 & 1991 29--30 Jul & 29.1\\
$rp23b$ &  0 & 1992 10 Aug & 5.0 \\
$rp23c$ &  0 & 1993 7--9 Jan   & 16.3\\
$rp07$ & 17 & 1993 5--5 Feb   & 3.7\\
$rp10$ & 14 & 1993 5--6 Feb   & 3.5\\
$rp89$ & 12 & 1993 6--7 Feb   & 4.5\\
\hline
\sc{hri} \hfill \\
\hline
$rh20a$ &  1 & 1992 8--12 Jan  & 19.1 \\
$rh20b$ &  1 & 1992 1--3 Aug   & 15.8 \\
$rh83$ & 14 & 1994 6--9 Aug   & 25.9 \\
$rh84$ & 13 & 1994 8--9 Aug   & 18.7 \\
$rh85$ & 16 & 1994 6--8 Aug   & 17.1 \\
$rh86$ &  7 & 1994 1--11 Aug  & 16.4 \\
$rh87$ & 17 & 1994 27 Jul--7 Aug  & 30.5 \\
$rh88$ & 13 & 1994 27 Jul--7 Aug  & 24.5 \\
$rh89$ & 10 & 1994 6--8 Aug   & 20.3 \\
$rh60a$ &  0 & 1994 10--11 Aug & 8.0  \\
$rh46$ &  0 & 1995 15 Jan & 24.6 \\
$rh60b$ &  0 & 1995 10--16 Jul & 40.9 \\
$rh11n$ &  0 & 1996 18 Jan--8 Feb  & 46.4 \\
$rh11a$ &  0 & 1996 17--27 Jul & 44.6 \\
$rh03$ &  0 & 1997 10--14 Jan & 33.9 \\
\hline
\end{tabular}
\tablenotetext{a}{The 
observation names correspond to the last digits of the {\it ROSAT} sequence 
numbers.}
\end{small}                                                                 
\end{center}
\label{tb:obs}
\end{table}

\pagebreak

\begin{figure}
\caption{Average corrected orbital fluxes for {\it rh8}. On-axis
observation {\it rh60a} can be seen at $d=947$. The $\chi^2$ test gives a 
$<$ 0.1\% probability that this is consistent with a constant flux.}
\label{fig:rh8}
\end{figure}

\begin{figure}
\caption{Long term lightcurve of X-8. Each point (HRI squares, PSPC circles)
 represents the average flux 
of each observation except at day$\approx$1000 which is the average of the 
off-axis observation + {\it rh60a}. The PSPC on-axis observations
were renormalized as explained in section 3. The origin is the 
start of {\it rh20a}. Dotted line is the average HRI flux.}
\label{fig:avg}
\end{figure}

\begin{figure}
\caption{CLEAN power spectrum of the $hri$ dataset (upper) and folded 
lightcurve of X-8 on 105.9 days (lower). The highest peak corresponds
to the 105.9 day period (confidence $>$99.8\%).
Each point in the folded data represents the average X-8 HRI flux during one 
{\it ROSAT} orbit. The renormalized PSPC data appear as circles at phases 
0.42 ({\it rp23a},{\it rp23c}) and 0.05 ({\it rp23b}). The renormalized 
mean fluxes for the {\it Einstein} HRI (diamonds) and IPC (triangles)
observation are also shown. The dotted line represents the average {\it ROSAT}
HRI flux. The data are shown twice for clarity.}
\label{fig:period}
\end{figure}

\begin{figure}
\plotone{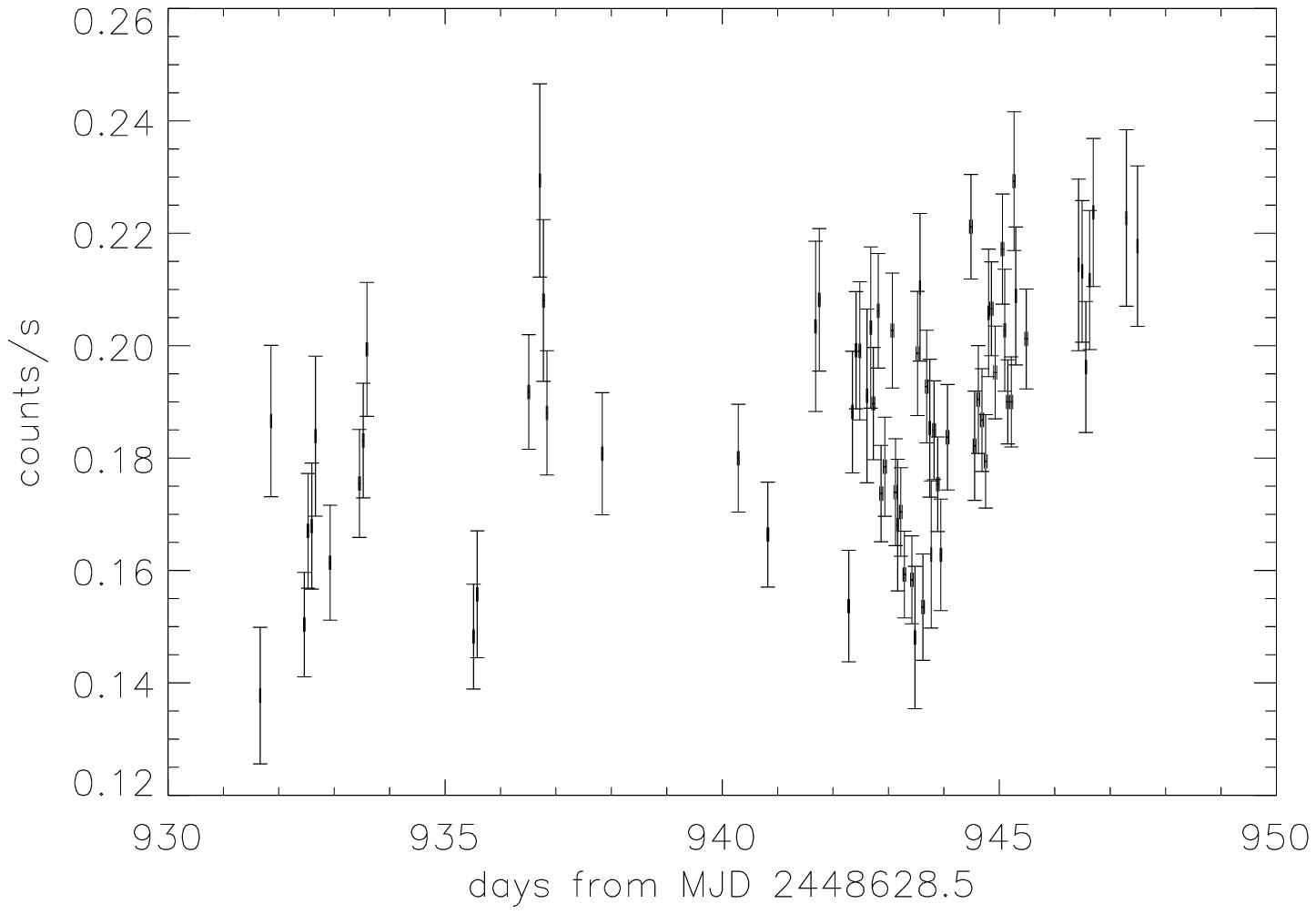}
\end{figure}

\begin{figure}
\plotone{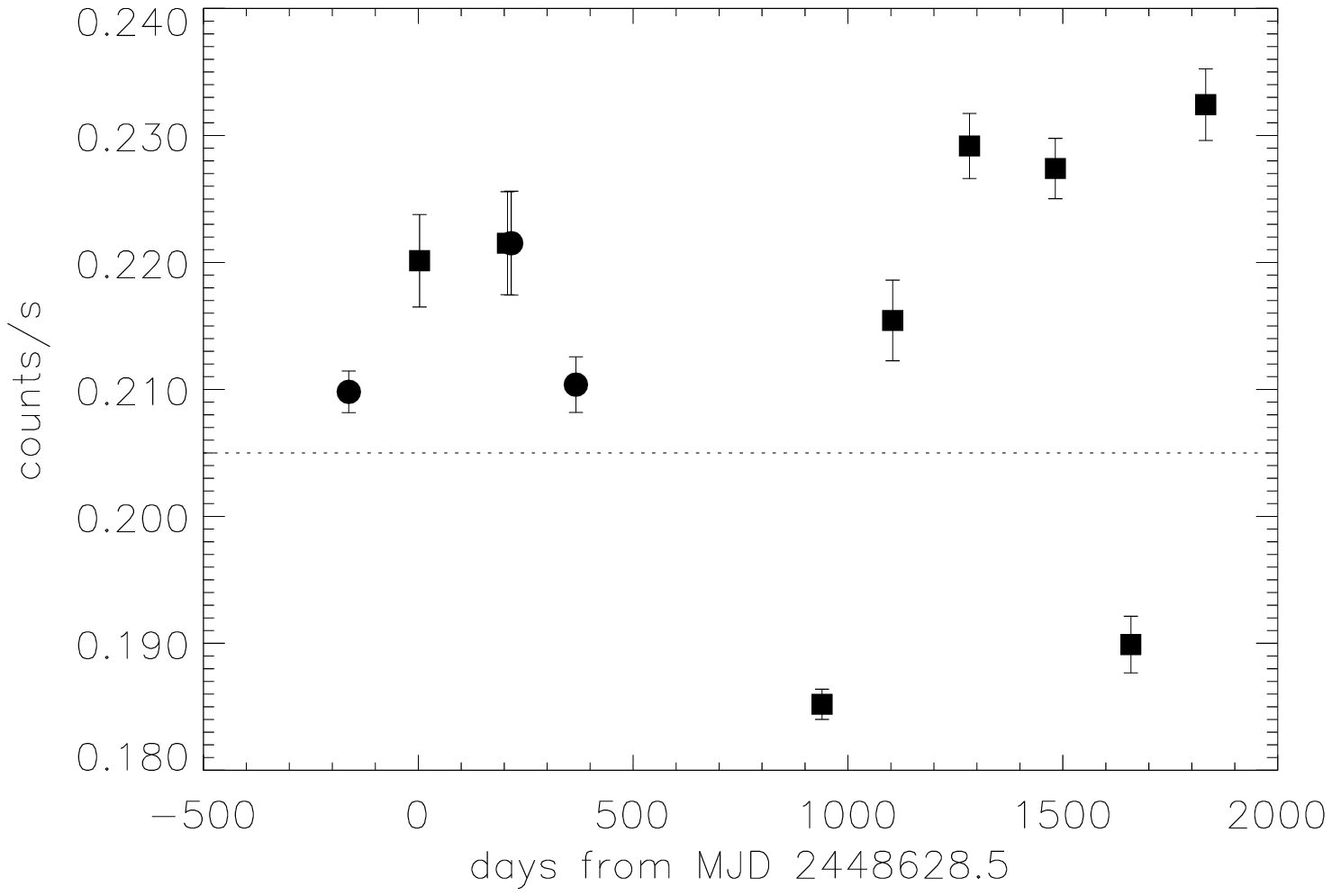}
\end{figure}

\begin{figure}
\plotone{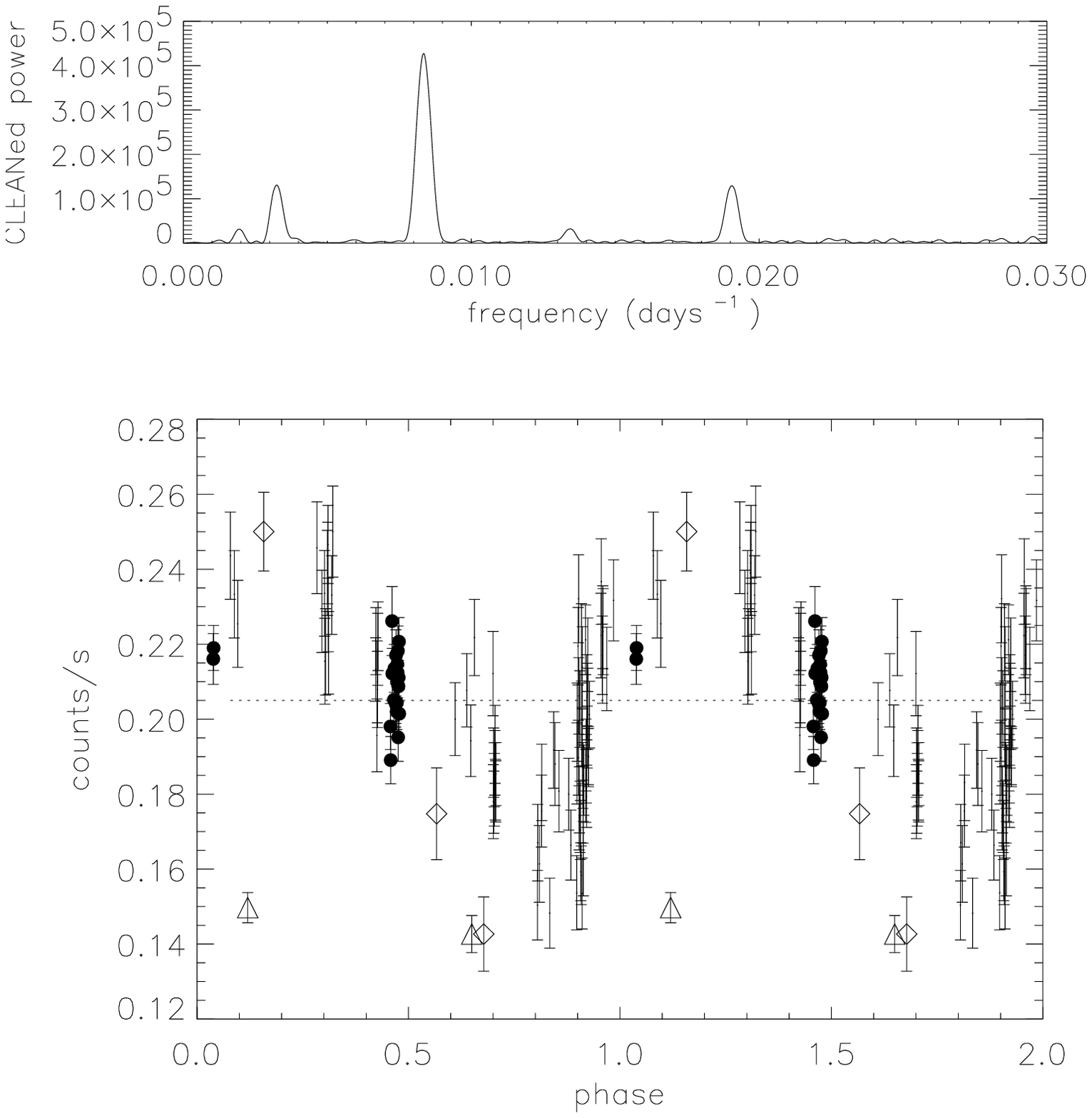}
\end{figure}
\end{document}